\documentclass[aps,prl,twocolumn,amsmath,amssymb]{revtex4-2}

\usepackage{graphicx, bm, xcolor, tikz}
\usepackage[colorlinks,allcolors=blue]{hyperref}
\usetikzlibrary{positioning, arrows.meta, shapes.geometric, fit}
\providecommand\apj{Astrophys. J.}
\providecommand\apjl{Astrophys. J. Lett.}
\providecommand\aap{Astron. Astrophys.}
\providecommand\mnras{Mon. Not. R. Astron. Soc.}

\providecommand\sovast{Sov. Astron.}
\providecommand\nat{Nature}
\providecommand\prl{Phys. Rev. Lett.}

\begin{document}

\title{The Birth of Plasmoids: Weibel-Driven Secondary Tearing and its Suppression in Relativistic Reconnection}

\author{Matthew Goodbred}
\email{mgoodbred@princeton.edu}
\author{Anatoly Spitkovsky}
\email{anatoly@astro.princeton.edu}
\affiliation{Department of Astrophysical Sciences, Princeton University, 4 Ivy Lane, Princeton, NJ 08544, USA}

\begin{abstract}

Plasmoids, or magnetic islands, dominate relativistic reconnection layers in kinetic simulations, yet the instability that creates them remains poorly understood. By tracking newborn islands in 2D particle-in-cell simulations of pair-plasma reconnection with magnetizations $0.1\le\sigma_0\le1000$, we show that plasmoids are seeded by a current-filamentation/Weibel instability driven by counter-streaming out-of-plane current carriers near primary x-lines. Measured growth rates are $\sim0.1$ times the background plasma frequency and agree with linear theory evaluated on the simulated distributions. As inflowing plasma loads the sheet away from the x-line, the current is carried by more particles and their out-of-plane drift slows. The filamentation instability shuts off once this drift becomes comparable to the momentum spread in the outflow direction, which occurs at a distance $\ell_{\rm iso}\propto\sigma_0$ from the x-line. Because the unstable region extends farther at higher $\sigma_0$, our results account for the onset of plasmoid-dominated reconnection in the relativistic regime but predict that plasmoid formation is suppressed beyond $\sim\ell_{\rm iso}$ from primary x-lines, even in ultra-relativistic reconnection.
Particle acceleration, variability, and coherent radio emission attributed to plasmoid-dominated reconnection may therefore be confined to smaller regions than previously assumed.

\end{abstract}

\maketitle

\section{Introduction}

By liberating free energy in sheared magnetic fields, magnetic reconnection heats plasma, accelerates particles, and drives outflows, potentially accounting for explosive and energetic phenomena in a range of astrophysical settings such as pulsar winds \citep[e.g.,][]{philippov_ab-initio_2018,cerutti_dissipation_2020, Hakobyan2023}, black hole coronae and jets \citep[e.g.,][]{Bransgrove2021,Mellah2023}, and magnetar magnetospheres \citep[e.g.,][]{yuan_plasmoid_2020,mahlmann_three-dimensional_2023}.  
Reconnection occurs at magnetic x-points (x-lines in 3D), where the magnetic field changes topology and bipolar outflows are launched. While non-relativistic reconnection can feature a single x-line and the broad outflow exhausts characterized by \citet{Petschek1964} \citep[][]{robbins_transition_2025}, kinetic simulations of relativistic reconnection ubiquitously show a more complicated geometry in which the reconnection layer breaks into a chain of O-points (islands) linked by thin current sheets containing x-points. In quasi-steady reconnection, the islands, also known as plasmoids, are continually ejected along the outflow, and the chain is replenished as the current sheets between them fragment to form new islands, a process known as the secondary tearing instability \citep[][]{Shibata2001,Uzdensky2010,Loureiro2012}.
Plasmoids then grow by accreting reconnected flux and merging, producing a hierarchical chain of a few large plasmoids separated by many small ones.
The chain contains two classes of x-lines, a distinction that will prove central to where plasmoids can form. \emph{Primary} x-lines are stationary in the lab frame and launch the large-scale bipolar outflows. Neighboring primary x-lines may be separated by large, quasi-stationary plasmoids. \emph{Advected} x-lines form between plasmoids that move with the bulk outflow.

Plasmoid chains have been recognized as key structures in the reconnection layer \citep[][]{sironi_plasmoids_2016,sironi_relativistic_2025}. They contribute significantly to particle acceleration through Fermi processes and mergers \citep[e.g.,][]{Drake2006,Sironi2014,Guo2015} and produce emission variability due to their non-uniform, hierarchical structure \citep[e.g.,][]{petropoulou_blazar_2016, Cerutti2012, Hosking2020}. 
Plasmoid mergers are invoked to explain coherent radio emission in pulsars and fast radio bursts (FRBs) \citep[e.g.,][]{Lyubarsky2019, Yyubarsky2020, mahlmann_electromagnetic_2022}.  

Despite their central importance in relativistic reconnection, the microphysics of why and where plasmoids form is not fully understood. 
The relativistic tearing instability has been studied extensively for an equilibrium Harris sheet \citep[e.g.,][]{Zelenzi1979,hoshino_stabilization_2020,Demidov2025}, but in active reconnection islands emerge from highly dynamic skin-depth scale current sheets with non-equilibrium particle distributions. In this work, we focus on the linear stage of plasmoid formation in such sheets. We show using 2D particle-in-cell (PIC) simulations that it proceeds through the current filamentation, or Weibel, instability \citep{Fried1959, Weibel1959}, driven by anisotropic current-carrying beams. Beyond identifying the physical origin of secondary tearing in kinetic-scale current sheets, this mechanism explains why relativistic reconnection appears plasmoid-dominated. We show why plasmoids grow near x-lines and predict that their formation is suppressed sufficiently far from the primary x-line as the sheet is loaded with plasma and the current drift velocity falls. 

\begin{figure*}[htb!]
    \centering
    \includegraphics[width=\linewidth]{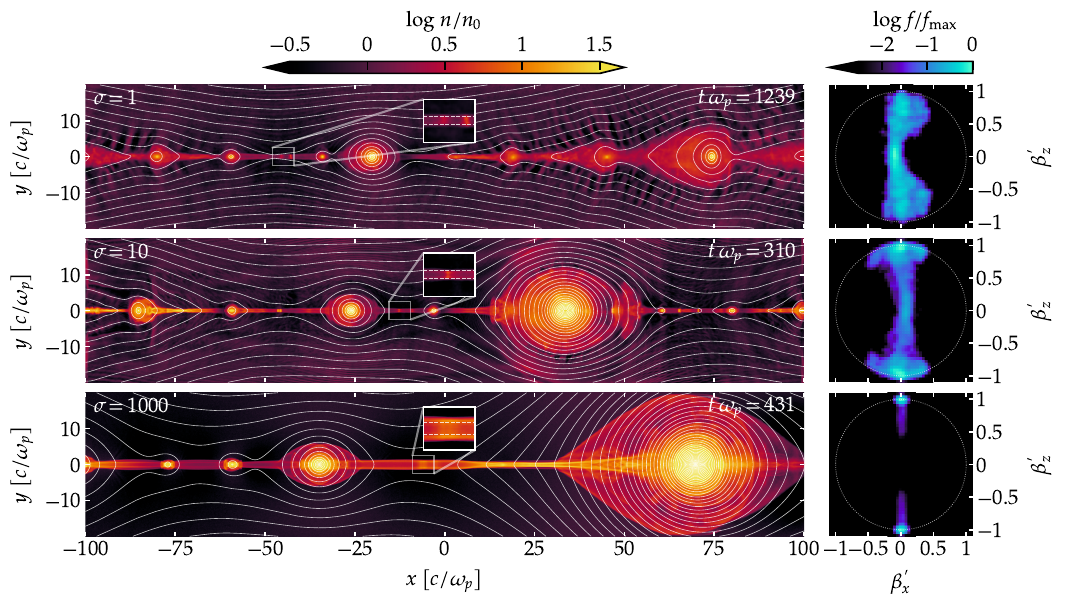}
    \caption{Left: Total plasma density with magnetic flux contours in white. Insets show examples of proto-plasmoid structures. $\sigma_0$ increases from top to bottom. Right: Particle histograms in $\beta_x'$ and $\beta_z'$, measured in the outflow rest frame, for particles of both species between the dashed white lines in the insets. Distributions are broad in $\beta_z'$ and narrow in $\beta_x'$, with anisotropy increasing with $\sigma_0$. }
    \label{fig:dens_dists}
\end{figure*}
\section{Simulations }\label{sec:sims}
We simulate 2D anti-parallel pair-plasma reconnection with the PIC code \textsc{Tristan-MP v2} \citep[][]{hayk_hakobyan_2023_7566725}. Inflow is along $y$, outflow along $x$, and $z$ is the out-of-plane direction. Each run is characterized by the upstream magnetization
\begin{equation}
    \sigma_0\equiv\frac{B_0^2}{4\pi n_0mc^2},
\end{equation}
where $B_0$ is the upstream reconnecting field and $n_0$ is the total upstream density. Since $\sigma_0$ is the available magnetic energy per particle in units of $mc^2$, $\sigma_0\gtrsim1$ marks the relativistic regime. We simulate $\sigma_0=0.1$, $1$, $10$, $50$, $250$, and $1000$.

The simulations are initialized using a Harris sheet, with upstream temperature $k_BT_0/mc^2=10^{-3}$ for all runs except $\sigma_0=1000$, with $k_BT_0/mc^2=10^{-1}$. A small sinusoidal perturbation to $B_y$ triggers reconnection at the center of the domain. Open boundaries in $x$ and inflow boundaries in $y$ allow the initial Harris sheet to be evacuated from the box, after which reconnection reaches a quasi-steady state. The upstream skin depth $d_0\equiv c/\omega_{p0}$, with $\omega_{p0}=(4\pi n_0e^2/m)^{1/2}$ the plasma frequency, is resolved by 20 cells. Each species is initialized with 25 particles per cell.

\paragraph{Notation.} For a particle with three-velocity $v_j$ ($j=x,y,z$), $\beta_j\equiv v_j/c$, $\gamma=(1-\beta^2)^{-1/2}$ is the Lorentz factor, and $u^\mu=(\gamma,\gamma\beta_j)$ is the four-velocity. Primed quantities are evaluated in the rest frame of the combined bulk flow of both species. Subscripts ``$0$" and ``${\rm cs}$'' denote far-upstream and current-sheet quantities, respectively. The uppercase $\Gamma$ refers to growth rates.

\section{X-line Anisotropy Drives Current Filamentation}\label{sec:anisotropy}
Near a primary x-line, the sheet must carry the current that reverses the upstream field, but it contains few particles to carry it. The current carriers therefore form fast, counter-streaming electron and positron beams along $z$, which is the classic configuration for the current filamentation, or Weibel, instability \citep{Weibel1959,Fried1959}. A small magnetic perturbation $\delta B_y$ deflects the counter-streaming, oppositely charged beams in the same direction along $x$, gathering parallel currents into filaments that reinforce the perturbation.
Since a thermal spread smears out bunching, the instability is strongest when the beams are cold in the direction of $\bm k=k\hat{\bm x}$, which is indeed the case near the x-line. The reconnection electric field $E_z$ accelerates inflowing particles out of the plane, focusing their momenta along $z$ \citep{Uzensky2011}, whereas $x$-momentum comes only from the $v_zB_y$ force that vanishes at the x-line \citep{Zenitani2001ThePlasmas}. The right panels of Fig.~\ref{fig:dens_dists} show the resulting beam-like distributions near x-lines for varying $\sigma_0$.

These beams drive the filamentation instability, which, in the linear stage, produces the small current and density enhancements that we call \textit{proto-plasmoids} (shown in the insets of Fig.~\ref{fig:dens_dists}). Once nonlinear, they become magnetic islands and, by accreting reconnected flux, grow into the plasmoids that dominate the sheet morphology in the main panels of Fig.~\ref{fig:dens_dists} \citep{sironi_plasmoids_2016}. 

To investigate the linear instability, we consider the transverse electromagnetic mode with ${\bm k}=k\,\hat{\bm x}$, $\delta {\bm B}=\delta B\,\hat{\bm y}$, and $\delta {\bm E}=\delta E\,\hat{\bm z}$. This is the polarization of the tearing mode, which is why the instability seeds magnetic islands. Since the magnetic field vanishes at the x-line, we take the background to be unmagnetized. The free energy of the sheared magnetic field that drives classical tearing is then absent by construction. Instead, the instability is driven entirely by momentum anisotropy. We also neglect the background $E_z$. Its effect on perturbed orbits is small given that the mode grows faster than particle ejection from the diffusion region, $\Gamma\tau_{\rm esc}\gtrsim 1$ with $\tau_{\rm esc}\sim E_{z}/(c\,\partial_xB_y)$, which we verify a posteriori.
Since wavelengths of growing modes are generally much shorter than the current sheet length, we use plane wave perturbations $\propto \exp(ikx+\Gamma t)$. 
The resulting dispersion relation, derived in Appendix~\ref{sec:appx_dispersion}, is 
\begin{equation}\label{eq:text_dispersion}
\begin{split}
    c^2k^2(1+M^2)=\omega_p^2\Bigg[&-\left\langle\frac{\gamma^2-u_z^2}{\gamma^3}\right\rangle-\left\langle\frac{2u_z^2u_x^2}{\gamma^3D}\right\rangle \\
    &+\left\langle\frac{u_z^2(\gamma^2-u_x^2)(\gamma^2 M^2-u_x^2)}{\gamma^3D^2}\right\rangle\Bigg],
\end{split}
\end{equation}
where $M\equiv \Gamma/ck$ is the imaginary phase velocity, $D\equiv u_x^2+\gamma^2 M^2$, and angle brackets denote averages over the equilibrium distribution. The momentum $u_y$ enters only through $\gamma$, so the instability is controlled by $u_x$ and $u_z$. The first two terms in Eq.~(\ref{eq:text_dispersion}) are stabilizing. The third, destabilizing term is weighted by $u_z^2$, revealing that the drive comes from the out-of-plane current beams. The same term is positive only for particles with $|\beta_x|<M$, i.e., $|v_x|<\Gamma/k$, so thermal spread along ${\bm k}$ stabilizes high-$k$ modes. 

\begin{figure}
    \centering
    \includegraphics[width=\linewidth]{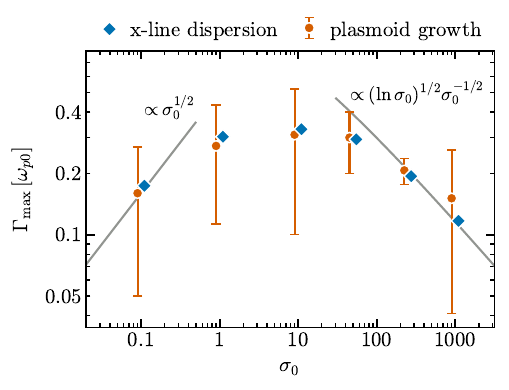}
    \caption{
    Maximum growth rates $\Gamma_{\rm max}$ of proto-plasmoids as a function of 
    $\sigma_0$. Blue diamonds are growth rates computed from the dispersion relation using distribution functions measured at x-lines in simulations. Orange circles are the measurements of average growth rates of linear current perturbations in the simulation current sheets, with 1-standard-deviation error bars. Markers are slightly offset along the x-axis for readability.
    }
    \label{fig:growth_rates}
\end{figure}

Using distributions measured at representative x-lines in the simulations (e.g., those highlighted in Fig.~\ref{fig:dens_dists}), we solve Eq.~(\ref{eq:text_dispersion}) for the growth rate $\Gamma_{\rm max}$ of the fastest-growing mode as a function of $\sigma_0$. The results, normalized to the upstream plasma frequency, are shown as blue diamonds in Fig.~\ref{fig:growth_rates}. 
Across our simulated range, $\Gamma_{\rm max}\sim0.1\,\omega_{p0}$, rising from small $\sigma_0$ to a peak near $\sigma_0\sim 10$ and falling at higher $\sigma_0$.

To test whether this instability drives the growth of proto-plasmoids,
we track midplane current perturbations at high cadence and fit their exponential growth rates (Appendix~\ref{sec:appx_tracker}). The mean measured rates, shown as orange circles in Fig.~\ref{fig:growth_rates}, agree well with the predictions of Eq.~(\ref{eq:text_dispersion}), confirming that the filamentation instability accounts for the exponential growth of linear fluctuations in kinetic-scale current sheets. 

The change in growth rates with $\sigma_0$ can be captured by simple scaling arguments. 
Neglecting the velocity spread in $\beta_x$ (see Appendix~\ref{sec:appx_dispersion}), the maximum growth rate is 
\begin{equation}
   \Gamma_{\rm max}=\omega_p\beta_J\left\langle\frac{1}{\gamma}\right\rangle^{1/2},
\end{equation}
where $\beta_J^2\equiv\langle\beta_z^2/\gamma\rangle/\langle1/\gamma\rangle$ is the inertia-weighted mean square drift velocity. 

For $\sigma_0\ll1$, $\gamma\approx 1$, and the sheet is nearly incompressible, so $\omega_p\approx\omega_{p0}$. If the current sheet half-width is equal to the local skin depth \citep{goodbred_first-principles_2022}, Amp\`ere's law gives $\beta_z\sim\sigma_0^{1/2}$, and hence $\Gamma_{\rm max}/\omega_{p0}\propto\sigma_0^{1/2}$.

In the $\sigma_0\gg 1$ limit, $\beta_J\approx 1$. The inflowing Poynting flux per particle gives $\langle\gamma\rangle\propto\sigma_0$. Pressure balance in the inflow direction implies $n_{\rm cs}\langle\gamma\rangle \sim n_0\sigma_0$, giving a current sheet density $n_{\rm cs}\sim n_0$. For an exponential energy distribution, $\langle1/\gamma\rangle\propto\sigma_0^{-1}\ln\sigma_0$, so $\Gamma_{\rm max}/\omega_{p0}\propto (\ln\sigma_0)^{1/2}\sigma_0^{-1/2}$. 
These scalings are shown in Fig.~\ref{fig:growth_rates}, confirming that they capture the limiting behavior of the growth rates. Growth rates rise with the drift velocity for $\sigma_0\ll 1$ and fall due to relativistic inertia for $\sigma_0\gg 1$.  

\section{Plasma loading shuts off filamentation}\label{sec:suppression}
Growth rates of order $0.1\,\omega_{p0}$ are fast compared to the local dynamical time, so perturbations can typically reach nonlinear amplitudes wherever the sheet is unstable. The regions where plasmoids grow are therefore set by the marginal stability boundary, the point at which counter-streaming 
beams no longer drive the instability. We now show that dilution of the current drift velocity due to loading by inflowing plasma sets this boundary.  

The strength of the counter-streaming beams that drive the instability follows from the availability of current carriers. 
Reversing the upstream field requires a current per unit length $\sim cB_0/4\pi$ that is independent of distance $\ell$ from the x-line. The number of particles available to carry it, however, grows with $\ell$, since inflow delivers plasma at a steady rate. The sheet's surface density therefore accumulates as $\Sigma\sim n_0\beta_{\rm rec}\ell(v_{A0}/v_{\rm out})$, where $v_{\rm in}$ and $v_{\rm out}$ are the local inflow and outflow speeds, respectively, $\beta_{\rm rec}\equiv v_{\rm in}/v_{A0}$, and $v_{A0}=[\sigma_0/(1+\sigma_0)]^{1/2}$ with the upstream Alfv\'en speed. Near the x-line, few particles carry the entire current and have large out-of-plane drift velocities, forming the counter-streaming beams that drive filamentation. 
Downstream, the same current is shared among more particles, and their drift falls as $\beta_z\propto v_{\rm out}/\ell$, which goes as $1/\ell$ once the outflow saturates at $v_{A0}$.

Meanwhile, each newly inflowing particle is picked up by the exhaust and gains an in-plane momentum spread comparable to the outflow speed. Filamentation shuts off where the spread of counter-streaming current carriers falls to this pickup spread. With increasing $\sigma_0$, the sheet must carry more current per particle, and its carriers have larger relativistic inertia, whereas the pickup spread is set by an outflow speed that saturates at $c$. The sheet must therefore accumulate more plasma before it becomes isotropic, extending the unstable region. We now quantify this physical picture.

As a proxy for stability to filamentation, we use the relativistic temperature ratio $\Theta'_x/\Theta'_z$, where $\Theta_i'=\langle u_i^2/\gamma\rangle$ is measured in the local outflow rest frame. Although exact only for a bi-Maxwellian (see Appendix~\ref{sec:appx_stabilization}), $\Theta'_x/\Theta'_z\ge 1$ is a useful indicator of local instability to filamentation for general distributions. 

Boosting the counter-streaming beams into the outflow frame gives an effective temperature $\Theta_z'\approx\gamma_{\rm out}\gamma\beta_z^2$. In this frame, inflowing particles enter the exhaust non-adiabatically and acquire an in-plane pickup temperature of $\Theta_x'\sim u_{\rm out}^2/\gamma_{\rm out}=\gamma_{\rm out}\beta_{\rm out}^2$ \citep{French2023}. The effective temperature ratio is then
\begin{equation}\label{eq:theta_ratio}
   \frac{\Theta_z'}{\Theta_x'}\approx\frac{\gamma\beta_z^2}{\beta_{\rm out}^2}.
\end{equation}
Inflowing Poynting flux supplies an energy per particle of $\gamma\approx 1+\sigma_0$. Finally, solving Amp\`ere's law for $\beta_z$ and eliminating the surface density using the continuity equation gives $\beta_z\approx \sigma_0^{1/2}(d_0/\beta_{\rm rec}\ell)(v_{\rm out}/v_{A0})$. Inserting these expressions into Eq.~(\ref{eq:theta_ratio}) yields
\begin{equation}\label{eq:anisotropy}
    \frac{\Theta_z'}{\Theta_x'}\approx(1+\sigma_0)^2\left(\frac{d_0}{\beta_{\rm rec}\ell}\right)^2.
\end{equation}
The anisotropy decays with $\ell^{-2}$ as the inflow loads the sheet with pairs and lowers the drift speed. Setting the ratio to unity gives the distance from the x-line at which the exhaust becomes isotropic,
\begin{equation}\label{eq:l_iso}
    \ell_{\rm iso}\sim\frac{1+\sigma_0}{\beta_{\rm rec}}\,d_0.
\end{equation}
$\beta_{\rm rec}\sim0.1$ independent of $\sigma_0$ \citep{liu_why_2017,goodbred_first-principles_2022}, so $\ell_{\rm iso}\propto\sigma_0$ for $\sigma_0\gg1$.

The distance $\ell$, and hence $\ell_{\rm iso}$, is measured from primary x-lines, not advected x-lines.
Outside $\ell_{\rm iso}$ from a primary x-line, the sheet is loaded with plasma, and advected x-lines cannot restore plasma-starved, unstable conditions which would reset $\ell$.
Their forward exhaust, aligned with the primary exhaust, cannot outrun the bulk flow once the trailing plasmoid is accelerated to $v_{\rm out}\sim c$ \citep{sironi_plasmoids_2016}, so reconnected plasma accumulates and dilutes the current drift velocity. Their reverse exhaust is blocked by the trailing plasmoid (e.g., the x-line at $\approx-60\,c/\omega_{p0}$ in the $\sigma_0=10$ panel of Fig.~\ref{fig:dens_dists}), which absorbs any filament seeded there. Unable to produce new plasmoids on either side, advected x-lines cannot extend plasmoid formation beyond $\sim\ell_{\rm iso}$ from a primary x-line. The exception is around rare, large plasmoids, which accelerate to $\sim c$ over distances that may be long compared to $\ell_{\rm iso}$. Until they reach $\sim c$, their leading x-line may still eject a few plasmoids into the primary outflow.

\begin{figure*}
    \centering
    \includegraphics[width=\linewidth]{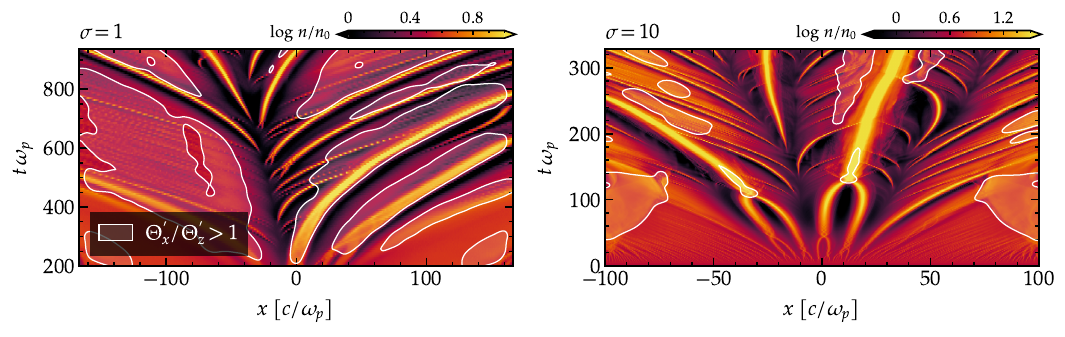}
    \caption{Time-stacks of midplane density, averaged over $2\,c/\omega_{p0}$ in the inflow direction, for $\sigma_0=1$ and $\sigma_0=10$. White shading marks where the temperature anisotropy parameter $\Theta_x'/\Theta_z'>1$, indicating stability to filamentation. 
    Axes are scaled so that the Alfv\'en speed has the same slope in both panels. Plasmoids are only born in the unshaded regions, which extend further from primary x-lines in the $\sigma_0=10$ case than $\sigma_0=1$. }
    \label{fig:plasmoid_timestack}
\end{figure*}

Figure~\ref{fig:plasmoid_timestack} shows density time-stacks for $\sigma_0=1$ and 10 runs. Plasmoids appear as overdensities that form spontaneously and propagate away from the underdense x-lines, while exhausts have intermediate density and are found between x-lines and the plasmoids or boundary. Shaded regions have $\Theta_x'/\Theta_z'>1$, indicating stability to filamentation. Plasmoids are born throughout the unshaded regions, anywhere within $\sim\ell_{\rm iso}$ of a primary x-line and outside existing plasmoids.

Plasmoid interiors are generally stable, either because scattering off in-plane fields isotropizes the distribution or because strong magnetization suppresses modes with wavelength larger than the local gyroradius. Exhausts become stable a distance $\sim\ell_{\rm iso}$ from primary x-lines. In the $\sigma_0=1$ run, isotropy is reached close to the x-lines and stable exhausts occupy most of the layer. In the $\sigma_0=10$ run, isotropy is reached several times farther out, and the layer is plasmoid-dominated. This is consistent with Eq.~(\ref{eq:l_iso}), which predicts $\ell_{\rm iso}(\sigma_0=10)/\ell_{\rm iso}(\sigma_0=1)\approx5.5$.

\section{Implications for Reconnection Morphology}\label{sec:morphology}

The restriction of plasmoid formation to within $\ell_{\rm iso}$ from primary x-lines accounts for $\sigma_0$-dependent changes in reconnection morphology. In the $\sigma_0\lesssim1$ regime, $\ell_{\rm iso}\approx d_0/\beta_{\rm rec}$, corresponding to the length of the diffusion region. Just outside, the thermal spread in the outflow direction is comparable to the spread of the current carriers. Therefore, filamentation occurs only very close to the x-line in non-relativistic reconnection. Plasmoid formation and mergers occur less frequently, enabling the development of open exhausts like in the top panel of Fig.~\ref{fig:dens_dists}.  

For $\sigma_0\gg1$, $\ell_{\rm iso}\approx\sigma_0d_0/\beta_{\rm rec}$, and plasmoids can form throughout anisotropic exhausts that extend far beyond the diffusion region.
Nonetheless, we predict that as the outflowing sheet accumulates plasma, the plasmoid formation that sustains the hierarchical chain is suppressed beyond a distance $\sim\ell_{\rm iso}$ from primary x-lines. 

This picture explains the results of \citet{robbins_transition_2025}, who found that rising plasmoid generation rates drive the change from Petschek-like layers at low $\sigma_0$ to plasmoid-dominated layers at high $\sigma_0$. 
Our results suggest that this reflects not an intrinsic property of relativistic reconnection, but the size of the simulation domain relative to a $\sigma_0$-dependent length, $\ell_{\rm iso}$. The reconnection layer should appear plasmoid-dominated for box sizes $L\lesssim\ell_{\rm iso}$ and give way to Petschek-like exhausts beyond $\ell_{\rm iso}$ when $L\gg\ell_{\rm iso}$. Since $\ell_{\rm iso}\propto\sigma_0$, extremely large domains are required to capture this transition for $\sigma_0\gg1$.  

\section{Conclusions}\label{sec:summary}

We have shown that the secondary tearing that generates plasmoids in anti-parallel pair-plasma reconnection is the current-filamentation, or Weibel, instability. Both the onset and the suppression of plasmoid formation follow from a single constraint: the sheet must carry the current that reverses the upstream field. Near primary x-lines, few particles are available to carry it, so the carriers form fast counter-streaming beams that are cold in the outflow direction and unstable to filamentation \citep{Weibel1959,Fried1959}. Across $0.1\le\sigma_0\le1000$, proto-plasmoids tracked in our simulations grow with rates $\sim0.1\,\omega_{p0}$, in agreement with linear theory evaluated on the simulated distributions.

Away from the x-line, inflow loads the sheet with plasma, and the same current is shared among more carriers. Their drift falls until the counter-streaming beams no longer dominate the in-plane momentum spread from pickup heating, and filamentation shuts off at $\ell_{\rm iso}\sim(1+\sigma_0)\,d_0/\beta_{\rm rec}$. Because the instability grows quickly, this stability boundary, not the growth rate, determines where plasmoids form. Although growth rates fall at high $\sigma_0$, the unstable, anisotropic region expands because each particle must carry more current and is heavier with relativistic inertia, while the thermal spread from pickup heating is set by the outflow speed which saturates at $c$.

Because $\ell_{\rm iso}$ grows with $\sigma_0$, this picture accounts for the onset of bursty, plasmoid-dominated reconnection in the relativistic regime. For $\sigma_0\lesssim1$, the unstable region is no larger than the diffusion region and the sheet has little room to fragment. For $\sigma_0\gg1$ unstable beams persist far into the exhaust, increasing the rate of plasmoid generation and opportunity for mergers.

Even for large $\sigma_0$, $\ell_{\rm iso}$ remains finite, approaching $\ell_{\rm iso}\sim10\,\sigma_0\,d_0$ with $\beta_{\rm rec}\sim0.1$. PIC domains rarely capture scales $\gg\ell_{\rm iso}$ at high $\sigma_0$, whereas astrophysical current sheets exceed $d_0$ by many orders of magnitude. The plasmoid-dominated sheets seen in simulations may therefore partly reflect their limited size.
Beyond $\ell_{\rm iso}$ from primary x-lines, existing plasmoids may still grow and merge, but the hierarchical chain cannot be replenished, and the sheet may relax to Petschek-like exhausts as in the non-relativistic regime \citep{robbins_transition_2025}. Phenomena attributed to plasmoid-dominated layers would then operate over smaller volumes than previously assumed. These include particle acceleration in coalescing plasmoids \citep[e.g.,][]{sironi_relativistic_2025}, coherent radio emission from plasmoid mergers proposed for Crab nanoshots \citep{Philippov2019_nanoshots}, pulsar giant pulses \citep{Lyubarsky2019}, and fast radio bursts \citep{Yyubarsky2020,mahlmann_electromagnetic_2022}, and variability from localized plasmoids \citep[e.g.,][]{petropoulou_blazar_2016,Cerutti2012,Hosking2020}.

Our linear analysis assumes a uniform, unmagnetized, stationary background. Including the reconnection electric field, bulk outflow, and reconnected field is a natural extension. Understanding the general case of an out-of-plane guide field is a crucial next step, since it should impose a low-$k$ cutoff where the gyroradius falls below the mode wavelength, consistent with reduced plasmoid generation in guide-field simulations \citep[e.g.,][]{ball_mechanism_2019,rowan_electron_2019,zhang_radiation_2020}. Strong radiative cooling should similarly limit plasmoid formation, since lowering the particle energy $\gamma$ reduces the anisotropy. Most directly, our results predict that in simulations with domain sizes $L\gg\ell_{\rm iso}$, plasmoid generation should cease beyond $\sim\ell_{\rm iso}$ from primary x-lines.

\begin{acknowledgments}
We are grateful to Adam Robbins for insightful discussions. This research was supported by the Multimessenger Plasma Physics Center (MPPC, NSF grants PHY-2206607) and by a grant from the Simons Foundation (MP-SCMPS-0000147).
\end{acknowledgments}

\appendix
\setcounter{secnumdepth}{2}
\section{Dispersion relation}\label{sec:appx_dispersion}
\subsection{Derivation}\label{sec:appx_derivation}
To obtain the Weibel instability dispersion relation, we linearize the Vlasov-Maxwell equations, allowing perturbed quantities (denoted with $\delta$) to vary as $\exp(ikx +\Gamma t)$ where $k$ is the wavenumber in the outflow direction and $\Gamma$ is the growth rate.
We consider the electromagnetic mode with $\delta{\bm E}=\delta E_z\hat{\bm z}$ oriented out of the reconnection plane and $\delta{\bm B}=\delta B_y\hat{\bm y}$. 
Since the magnetic field vanishes at the x-line of antiparallel reconnection, we use an unmagnetized background. The unperturbed distribution function $f_s^0({\bm u})$ is uniform and normalized to the species density, $\int f_s^0\,d^3u=n_s$.  The distributions are anti-symmetric between species in $u_z$, since charges carry current in the z-direction, and symmetric in all other coordinates.
After linearizing and applying Faraday's law to eliminate $\delta E$, we obtain the perturbation to the distribution function for a single species:
\begin{equation}\label{eq:delta_f}
  \delta f_s=\frac{q_s}{m_s c^2}\delta B\left[\frac{i}{k}\frac{\partial f_s^0}{\partial u_z}+\frac{v_z}{\Gamma+ik v_x}\frac{\partial f_s^0}{\partial u_x}\right].
\end{equation}
The linearized Amp\`ere's law reads 
\begin{equation}\label{eq:delta_ampere}
    k^2(1+M^2)\delta B=-\frac{4\pi}{c} i k\,\delta J_z, 
\end{equation}
where we have introduced $M=\Gamma/kc$. Summing over species $s$, the current perturbation is 
\begin{equation}\label{eq:delta_J}
    \delta J_z=\sum_s\int q_s\,\delta f_s \,v_z\,d^3u.
\end{equation}
Combining Eqs.~(\ref{eq:delta_f}, \ref{eq:delta_ampere}, \ref{eq:delta_J}) yields 
\begin{equation}
\begin{split}
    c^2k^2(1+M^2)=\sum_s\frac{4\pi q_s^2}{m_sc}\int\Bigg[&v_z\frac{\partial f_s^0}{\partial u_z}\\
    &-\frac{v_z^2}{v_x-icM}\frac{\partial f_s^0}{\partial u_x}\Bigg]d^3u.
\end{split}
\end{equation}
The integrand is even in $v_z$ and therefore has equal contribution from both species. Writing $v_j/c=\beta_j$ and integrating by parts to eliminate derivatives of $f^0$, we obtain
\begin{equation}\label{eq:disp_unrationalized}
\begin{split}
    c^2k^2(1+M^2)=-\omega_p^2\Bigg[&\left\langle\frac{\gamma^2-u_z^2}{\gamma^3}\right\rangle+\left\langle\frac{2u_z^2u_x}{\gamma^4(\beta_x-iM)}\right\rangle \\
    &+\left\langle\frac{\beta_z^2(\gamma^2-u_x^2)}{\gamma^3(\beta_x-iM)^2}\right\rangle\Bigg],
\end{split}
\end{equation}
where $\omega_p^2=8\pi n_s q^2/m$ is the local plasma frequency of both species and the angle brackets denote averages: $\langle Q\rangle=n_s^{-1}\int Q\, f_s^0\,d^3u$. Assuming that $f^0$ is even in $u_x$ allows us to drop terms odd in $u_x$. Letting $\Gamma$ be real, further algebra to rationalize the denominators results in 
\begin{equation}\label{eq:appx_disp_raltionalized}
\begin{split}
    c^2k^2(1+M^2)=\omega_p^2\Bigg[&-\left\langle\frac{\gamma^2-u_z^2}{\gamma^3}\right\rangle-\left\langle\frac{2u_z^2u_x^2}{\gamma^3D}\right\rangle \\
    &+\left\langle\frac{u_z^2(\gamma^2-u_x^2)(\gamma^2 M^2-u_x^2)}{\gamma^3D^2}\right\rangle\Bigg],
\end{split}
\end{equation}
where $D=u_x^2+\gamma^2 M^2$. Now assume a Gaussian, nonrelativistic distribution in $\beta_x$, giving $f_s^0=\gamma n_s F(u_y,u_z)g(\beta_x)$ with $g(\beta_x)=\exp(-\beta_x^2/2\theta_x)/\sqrt{2\pi\theta_x}$ and $\theta_x\ll 1$.  Then we can further simplify Eq.~(\ref{eq:appx_disp_raltionalized}). Since $\gamma^2\approx 1 + u_y^2+u_z^2 $ in the small $\beta_x$ limit, we neglect the contribution of $u_x$ to $\gamma$. Then, 
\begin{equation}\label{eq:appx_dist_simplify1}
\begin{split}
    c^2k^2(1+M^2)=\omega_p^2\Bigg[&-\left\langle\frac{\gamma^2-u_z^2}{\gamma^3}\right\rangle-\left\langle\frac{2u_z^2}{\gamma^3}\frac{\beta_x^2}{ M^2+\beta_x^2}\right\rangle \\
    &+\left\langle\frac{u_z^2}{\gamma^3}\frac{ M^2-\beta_x^2}{(M^2+\beta_x^2)^2}\right\rangle\Bigg].
\end{split}
\end{equation}
Start by taking the $\beta_x$ average of the final term. With ${(M^2-\beta_x^2)/(M^2+\beta_x^2)^2}=(d/d\beta_x)[\beta_x/(M^2+\beta_x^2)]$, we can integrate by parts to obtain 
\begin{equation}
    \left\langle\frac{M^2-\beta_x^2}{(M^2+\beta_x^2)^2}\right\rangle_{\beta_x}=\frac{1}{\theta_x}\left\langle\frac{\beta_x^2}{M^2+\beta_x^2}\right\rangle_{\beta_x}.
\end{equation}
Therefore the second term on the right-hand side of Eq.~(\ref{eq:appx_dist_simplify1}) is smaller than the third by a factor of $\theta_x\ll 1$, and we can neglect the second term. Writing $\beta_x^2/(M^2+\beta_x^2)=1-M^2/(M^2+\beta_x^2)$, we obtain
\begin{align}
    \left\langle\frac{\beta_x^2}{ M^2+\beta_x^2}\right\rangle_{\beta_x}&=1-\frac{\zeta^2}{\sqrt{\pi}}\int\frac{e^{-t^2}}{t^2+\zeta^2}dt\\\notag
    &=1-\sqrt{\pi}\zeta e^{\zeta^2}{\rm erfc}(\zeta)\\\notag
    &\equiv W(\zeta)
\end{align}
where $\zeta= M/\sqrt{2\theta_x}$ and ${\rm erfc}(\zeta)=1-{\rm erf}(\zeta)$ is the complementary error function. Inserting this result in Eq.~(\ref{eq:appx_dist_simplify1}), 
\begin{equation}\label{eq:appx_dist_simplify2}
    c^2k^2(1+M^2)=\omega_p^2\left[\frac{W(\zeta)}{\theta_x}\left\langle\frac{\beta_z^2}{\gamma}\right\rangle -\left\langle\frac{1-\beta_z^2}{\gamma}\right\rangle\right].
\end{equation}
Define $\beta_J^2\equiv\langle\beta_z^2/\gamma\rangle/\langle1/\gamma\rangle$, the inertia-weighted mean-square drift velocity, and $\gamma_J^2=1/(1-\beta_J^2)$. Writing $ck=\Gamma/\zeta\sqrt{2\theta_x}$, the largest growth rate can be found by maximizing over the dimensionless parameter $\zeta$:
\begin{gather}
    \Gamma_{\rm max}=\omega_p\beta_J\left\langle\frac{1}{\gamma}\right\rangle^{1/2}\eta,\notag\\
    \eta^2=\underset{\zeta}{{\rm max}}\left[\frac{2\zeta^2}{1+2\theta_x\zeta^2}\left(W(\zeta)-\frac{\theta_x}{\gamma_J^2\beta_J^2}\right)\right],
\end{gather}
For cold beams with $\theta_x=0$, $\eta\rightarrow 1$, and 
\begin{equation}\label{eq:appx_cold_growth_rate}
   \Gamma_{\rm max}=\omega_p\beta_J\left\langle\frac{1}{\gamma}\right\rangle^{1/2},
\end{equation}
achieved in the limit of $k\rightarrow \infty$. 

\subsection{Stabilization criteria}\label{sec:appx_stabilization}
To find the criteria for stabilization of all modes, take the marginal stability limit $M\rightarrow 0$ of Eq.~(\ref{eq:appx_disp_raltionalized}). The first two terms in the bracket simplify to $-\langle(\gamma^2+u_z^2)/\gamma^3\rangle$. The third term must be regularized, since it diverges when $M=0$ and $\beta_x=0$. To do so, first use $\gamma_\perp^2=\gamma^2-u_x^2$ to write
\begin{equation}
\begin{split}
   &\lim_{M\rightarrow 0} \left\langle\frac{u_z^2(\gamma^2-u_x^2)(\gamma^2 M^2-u_x^2)}{\gamma^3D^2}\right\rangle\\
   &\quad=  \lim_{M\rightarrow 0} \left\langle\frac{u_z^2\gamma_\perp^2}{\gamma^5}\frac{M^2-\beta_x^2}{(\beta_x^2+M^2)^2}\right\rangle.
\end{split}
\end{equation}
Noting that ${M^2-\beta_x^2}/{(\beta_x^2+M^2)^2}=\partial_{\beta_x}(\beta_x/[\beta_x^2+M^2])$ and using $du_x/d\beta_x=\gamma^3/\gamma_\perp^2$, the average becomes 
\begin{equation}
    \frac{1}{n_s}\int du_y\,du_z\,d\beta_x\,f_s^0\frac{u_z^2}{\gamma^2}\frac{\partial}{\partial\beta_x}\left(\frac{\beta_x}{\beta_x^2+M^2}\right).
\end{equation}
Integrating by parts and switching back to integration over $u_x$, we have 
\begin{equation}
\begin{split}
    &-\frac{1}{n_s}\int d^3u\,\frac{\gamma u_xu_z^2}{(u_x^2+\gamma^2M^2)}\frac{\partial}{\partial u_x}\left(\frac{f_s^0}{\gamma^2}\right)\\
    &\quad\underset{M\rightarrow 0}{\rightarrow}   -\frac{1}{n_s}\int d^3u\,\frac{\gamma u_z^2}{u_x}\frac{\partial}{\partial u_x}\left(\frac{f_s^0}{\gamma^2}\right).
\end{split}
\end{equation}
Integrating by parts with $\gamma\partial_{u_x}(f_s^0/\gamma^2)\to\partial_{u_x}(f_s^0/\gamma)-f_s^0u_x/\gamma^3$, we obtain
\begin{equation}
\begin{split}
     &\lim_{M\rightarrow 0} \left\langle\frac{u_z^2(\gamma^2-u_x^2)(\gamma^2 M^2-u_x^2)}{\gamma^3D^2}\right\rangle\\
     &\quad=\left\langle\frac{u_z^2}{\gamma^3}\right\rangle-\frac{1}{n_s}\int d^3u\,\frac{u_z^2}{u_x}\frac{\partial}{\partial u_x}\left(\frac{f_s^0}{\gamma}\right).
\end{split}
\end{equation}
The dispersion relation in the $M\rightarrow 0$ limit then becomes
\begin{equation}
    \frac{c^2k^2}{\omega_p^2}=-\left\langle\frac{1}{\gamma}\right\rangle-\frac{1}{n_s}\int d^3u\,\frac{u_z^2}{u_x}\frac{\partial}{\partial u_x}\left(\frac{f_s^0}{\gamma}\right). 
\end{equation}
For the growth rate to vanish, $k\rightarrow 0 $ since $k=\Gamma/cM$. Therefore, marginal stability is attained when 
\begin{align}
  & \frac{1}{n_s}\int d^3u\,\frac{u_z^2}{u_x}\frac{\partial}{\partial u_x}\left(\frac{f_s^0}{\gamma}\right)+ \left\langle\frac{1}{\gamma}\right\rangle=0\\
   \implies &\int d^3u\,\left(\frac{u_z^2-u_x^2}{u_x}\right)\frac{\partial}{\partial  u_x}\left(\frac{f_s^0}{\gamma}\right)=0,
\end{align}
where the second line follows from integrating $\langle1/\gamma\rangle$ by parts and dropping the term that is odd in $u_x$. Expanding the derivative using the product rule results in 
\begin{equation}
\int d^3u\,\left[\left(\frac{u_z^2-u_x^2}{\gamma^3}\right)f_s^0-\left(\frac{u_z^2-u_x^2}{\gamma u_x}\right)\frac{\partial f_s^0}{\partial  u_x}\right]=0.
\end{equation}
With $(u_z^2-u_x^2)f_s^0/\gamma^3=(u_z/\gamma)\partial_{u_z}f_s^0-(u_x/\gamma)\partial_{u_x}f_s^0$ and $\partial_{u_i}=2u_i\partial_{u_i^2}$, this becomes
\begin{equation}\label{eq:appx_stabilization_criteria}
2\int d^3u\,\frac{u_z^2}{\gamma}\left[\frac{\partial f_s^0}{\partial (u_z^2)}-\frac{\partial f_s^0}{\partial  (u_x^2)}\right]=0.
\end{equation}
This form makes the physical picture clear. $\partial f_s^0/\partial u_j^2$ quantifies the local spread of the momentum distribution. For Gaussian-distributed momenta, as in a multi-temperature Maxwellian, $2\partial f_s^0/\partial u_j^2=-f_s^0/\theta_j$,  the inverse of the temperature in that direction, and all modes are stabilized when $\theta_z=\theta_x$. 

\section{Plasmoid tracker}\label{sec:appx_tracker}

Proto-plasmoids in nonlinear reconnecting current sheets do not emerge as Fourier modes across the length of the sheet, as in the tearing mode of an equilibrium current sheet. Instead, proto-plasmoids begin as clusters of one or a few discrete current perturbations in the secondary current sheets between larger plasmoids. In addition, the current sheets are highly dynamic and contain significant spatial structure. For those reasons, we do not measure growth rates by Fourier mode amplitudes but instead by tracking the linear proto-plasmoid fluctuations across timesteps. 

\begin{figure}
\centering
\begin{tikzpicture}[
    node distance = 3.5mm,
    box/.style = {draw, rounded corners=2pt, text width=64mm, align=center,
                  inner xsep=4pt, inner ysep=2.5pt, font=\small},
    sub/.style = {font=\footnotesize},
    >={Stealth[length=2mm]},
  ]
  \node[box] (det) {\textbf{Detect candidates}\\[1pt]
    {\footnotesize local maxima with $0.07<\delta J<1.5$, spacing
     $\ge 0.5\,c/\omega_{p0}$, $\delta n$ peak within $\pm2$ cells}};
  \node[box, below=of det] (lnk) {\textbf{Link to tracks}\\[1pt]
   {\footnotesize predict by linear extrapolation; require $|\Delta x|<v_{A0}\Delta t$;
    Hungarian match (min $\sum|\Delta x|$); unmatched $\to$ new track}
     };
  \node[box, below=of lnk] (trm) {\textbf{Terminate tracks}\\[1pt]
    {\footnotesize $\delta J>1.5$; unmatched $>2$ steps;
     merger within $2\,c/\omega_{p0}$}};
  \coordinate (r) at ([xshift=6mm]lnk.east);
  \node[draw, dashed, rounded corners, inner xsep=7pt, inner ysep=5pt,
        fit=(det)(lnk)(trm)(r)] (loopbox) {};
  \node[sub, anchor=south west] at (loopbox.north west)
        {for each output time $t$};
  \draw[->] (det) -- (lnk);
  \draw[->] (lnk) -- (trm);
  \draw[->] (trm.east) -- ++(6mm,0) |- (det.east);
  \node[box, below=6mm of loopbox] (sel) {\textbf{Select tracks}\\[1pt]
    {\footnotesize $\ge3$ points, rate $>0.05\,\omega_{p0}$; after initial
     Harris sheet; $\ge2$ points before $\delta J>1$ or decline}};
  \node[box, below=of sel] (gr) {\textbf{Fit growth rate}\\[1pt]
    {\footnotesize slope of $\ln\delta J(t)$ until $\delta J>1$ or 2
     decreases; drop max/min if $>5$ rates}};
  \draw[->] (loopbox.south) -- node[right, sub]{$t = t_{\rm end}$} (sel);
  \draw[->] (sel) -- (gr);
\end{tikzpicture}
\caption{Plasmoid-tracking algorithm. Steps inside the dashed box are repeated at
every output time; the remaining steps are applied once to the finished tracks.}
\label{fig:tracker}
\end{figure}
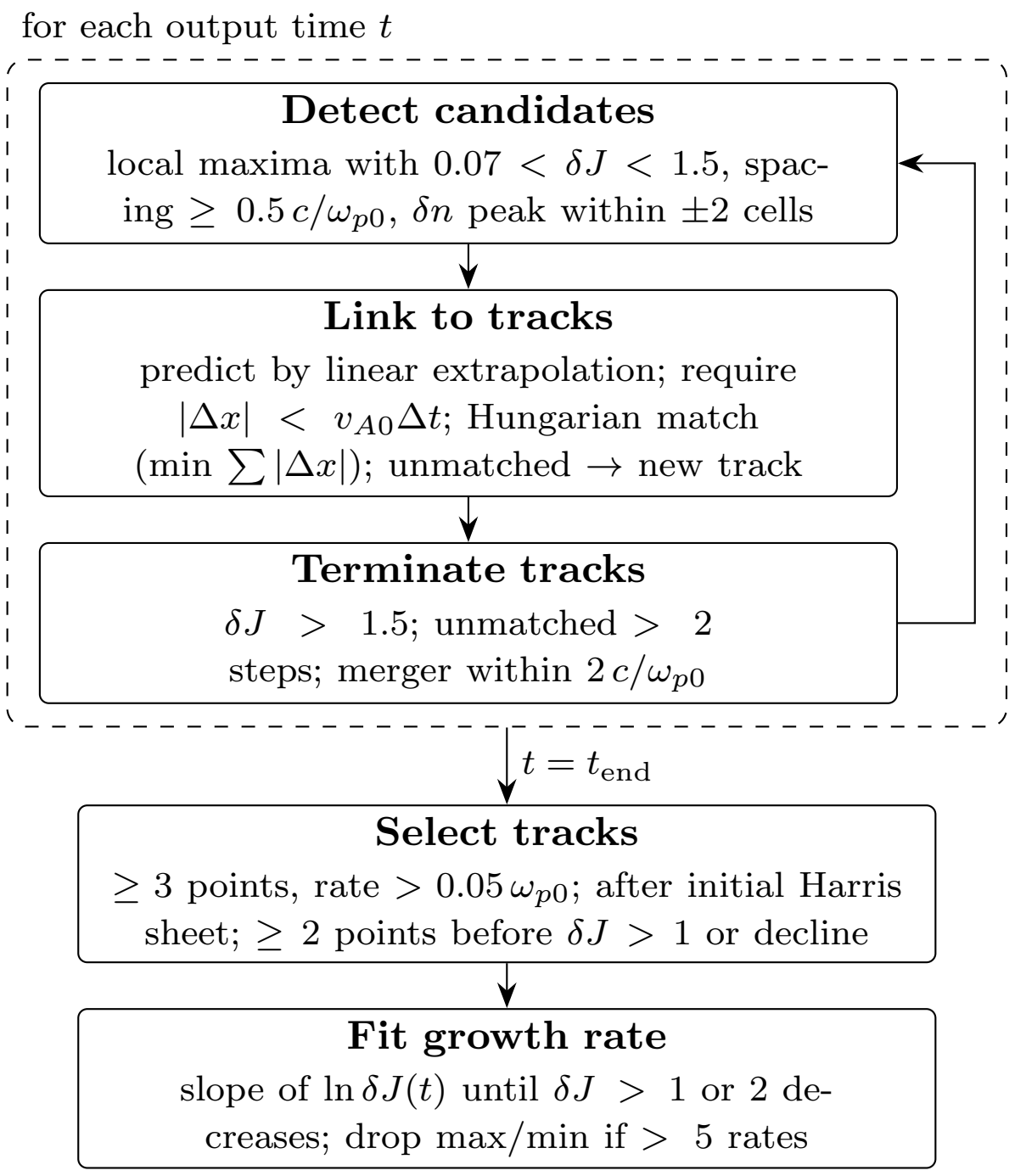

At each output time we use
\begin{equation}
  \delta J(x) = \frac{J_z - J_{z,\rm bg}}{J_{z,\rm bg}},
\end{equation}
where $J_z(x)$ is averaged over $|y| < 0.5\,c/\omega_{p0}$ and $J_{z,\rm bg}$ is $J_z$
smoothed with a Gaussian of width 20 cells. The density perturbation $\delta n(x)$ is
defined in the same way. The tracker, diagrammed in Fig.~\ref{fig:tracker}, proceeds as follows.

\begin{enumerate}
  \item \textbf{Detect plasmoid candidates.} Candidates are local maxima of $\delta J$ with
        $0.07 < \delta J < 1.5$, separated by at least $0.5\,c/\omega_{p0}$, with a
        $\delta n$ peak within $\pm 2$ cells.
  \item \textbf{Link candidates to tracks.} Each active track's position is predicted by linear
        extrapolation from its last two points (its last point if it has only one).
        Candidates within $v_{A0}\Delta t$ of a prediction are assigned to tracks by
        the Hungarian algorithm, minimizing $\sum|\Delta x|$. Unmatched candidates
        start new tracks.
  \item \textbf{Terminate tracks.} A track stops being followed if
        (a) $\delta J > 1.5$,
        (b) it goes unmatched for more than 2 consecutive steps, or
        (c) it comes within $2\,c/\omega_{p0}$ of another track and has fewer points (i.e., the tracks merge).
  \item \textbf{Select final tracks.} After the run, keep tracks that
        (a) have $\ge 3$ points and a growth rate $> 0.05\,\omega_{p0}$,
        (b) begin after the initial Harris sheet has formed plasmoids or left the
        domain, and
        (c) have $\ge 2$ points before $\delta J > 1$ or $\delta J$ starts to fall.
  \item \textbf{Fit growth rates.} For each track, the growth rate is the slope of a linear fit
        to $\ln \delta J(t)$ from the start of the track until $\delta J > 1$ or
        $\delta J$ falls for 2 consecutive steps. If more than 5 tracks remain, the
        highest and lowest rates are discarded.
\end{enumerate}

\bibliographystyle{apsrev4-2}
\bibliography{paper}

\begin{thebibliography}{38}%
\makeatletter
\providecommand \@ifxundefined [1]{%
 \@ifx{#1\undefined}
}%
\providecommand \@ifnum [1]{%
 \ifnum #1\expandafter \@firstoftwo
 \else \expandafter \@secondoftwo
 \fi
}%
\providecommand \@ifx [1]{%
 \ifx #1\expandafter \@firstoftwo
 \else \expandafter \@secondoftwo
 \fi
}%
\providecommand \natexlab [1]{#1}%
\providecommand \enquote  [1]{``#1''}%
\providecommand \bibnamefont  [1]{#1}%
\providecommand \bibfnamefont [1]{#1}%
\providecommand \citenamefont [1]{#1}%
\providecommand \href@noop [0]{\@secondoftwo}%
\providecommand \href [0]{\begingroup \@sanitize@url \@href}%
\providecommand \@href[1]{\@@startlink{#1}\@@href}%
\providecommand \@@href[1]{\endgroup#1\@@endlink}%
\providecommand \@sanitize@url [0]{\catcode `\\12\catcode `\$12\catcode `\&12\catcode `\#12\catcode `\^12\catcode `\_12\catcode `\%12\relax}%
\providecommand \@@startlink[1]{}%
\providecommand \@@endlink[0]{}%
\providecommand \url  [0]{\begingroup\@sanitize@url \@url }%
\providecommand \@url [1]{\endgroup\@href {#1}{\urlprefix }}%
\providecommand \urlprefix  [0]{URL }%
\providecommand \Eprint [0]{\href }%
\providecommand \doibase [0]{https://doi.org/}%
\providecommand \selectlanguage [0]{\@gobble}%
\providecommand \bibinfo  [0]{\@secondoftwo}%
\providecommand \bibfield  [0]{\@secondoftwo}%
\providecommand \translation [1]{[#1]}%
\providecommand \BibitemOpen [0]{}%
\providecommand \bibitemStop [0]{}%
\providecommand \bibitemNoStop [0]{.\EOS\space}%
\providecommand \EOS [0]{\spacefactor3000\relax}%
\providecommand \BibitemShut  [1]{\csname bibitem#1\endcsname}%
\let\auto@bib@innerbib\@empty
%</preamble>
\bibitem [{\citenamefont {Philippov}\ and\ \citenamefont {Spitkovsky}(2018)}]{philippov_ab-initio_2018}%
  \BibitemOpen
  \bibfield  {author} {\bibinfo {author} {\bibfnamefont {A.~A.}\ \bibnamefont {Philippov}}\ and\ \bibinfo {author} {\bibfnamefont {A.}~\bibnamefont {Spitkovsky}},\ }\href {https://doi.org/10.3847/1538-4357/aaabbc} {\bibfield  {journal} {\bibinfo  {journal} {The Astrophysical Journal}\ }\textbf {\bibinfo {volume} {855}},\ \bibinfo {pages} {94} (\bibinfo {year} {2018})}\BibitemShut {NoStop}%
\bibitem [{\citenamefont {Cerutti}\ \emph {et~al.}(2020)\citenamefont {Cerutti}, \citenamefont {Philippov},\ and\ \citenamefont {Dubus}}]{cerutti_dissipation_2020}%
  \BibitemOpen
  \bibfield  {author} {\bibinfo {author} {\bibfnamefont {B.}~\bibnamefont {Cerutti}}, \bibinfo {author} {\bibfnamefont {A.~A.}\ \bibnamefont {Philippov}},\ and\ \bibinfo {author} {\bibfnamefont {G.}~\bibnamefont {Dubus}},\ }\href {https://doi.org/10.1051/0004-6361/202038618} {\bibfield  {journal} {\bibinfo  {journal} {Astronomy \& Astrophysics}\ }\textbf {\bibinfo {volume} {642}},\ \bibinfo {pages} {A204} (\bibinfo {year} {2020})}\BibitemShut {NoStop}%
\bibitem [{\citenamefont {{Hakobyan}}\ \emph {et~al.}(2023)\citenamefont {{Hakobyan}}, \citenamefont {{Philippov}},\ and\ \citenamefont {{Spitkovsky}}}]{Hakobyan2023}%
  \BibitemOpen
  \bibfield  {author} {\bibinfo {author} {\bibfnamefont {H.}~\bibnamefont {{Hakobyan}}}, \bibinfo {author} {\bibfnamefont {A.}~\bibnamefont {{Philippov}}},\ and\ \bibinfo {author} {\bibfnamefont {A.}~\bibnamefont {{Spitkovsky}}},\ }\href {https://doi.org/10.3847/1538-4357/acab05} {\bibfield  {journal} {\bibinfo  {journal} {\apj}\ }\textbf {\bibinfo {volume} {943}},\ \bibinfo {eid} {105} (\bibinfo {year} {2023})},\ \Eprint {https://arxiv.org/abs/2209.02121} {arXiv:2209.02121 [astro-ph.HE]} \BibitemShut {NoStop}%
\bibitem [{\citenamefont {{Bransgrove}}\ \emph {et~al.}(2021)\citenamefont {{Bransgrove}}, \citenamefont {{Ripperda}},\ and\ \citenamefont {{Philippov}}}]{Bransgrove2021}%
  \BibitemOpen
  \bibfield  {author} {\bibinfo {author} {\bibfnamefont {A.}~\bibnamefont {{Bransgrove}}}, \bibinfo {author} {\bibfnamefont {B.}~\bibnamefont {{Ripperda}}},\ and\ \bibinfo {author} {\bibfnamefont {A.}~\bibnamefont {{Philippov}}},\ }\href {https://doi.org/10.1103/PhysRevLett.127.055101} {\bibfield  {journal} {\bibinfo  {journal} {\prl}\ }\textbf {\bibinfo {volume} {127}},\ \bibinfo {eid} {055101} (\bibinfo {year} {2021})},\ \Eprint {https://arxiv.org/abs/2109.14620} {arXiv:2109.14620 [astro-ph.HE]} \BibitemShut {NoStop}%
\bibitem [{\citenamefont {{El Mellah}}\ \emph {et~al.}(2023)\citenamefont {{El Mellah}}, \citenamefont {{Cerutti}},\ and\ \citenamefont {{Crinquand}}}]{Mellah2023}%
  \BibitemOpen
  \bibfield  {author} {\bibinfo {author} {\bibfnamefont {I.}~\bibnamefont {{El Mellah}}}, \bibinfo {author} {\bibfnamefont {B.}~\bibnamefont {{Cerutti}}},\ and\ \bibinfo {author} {\bibfnamefont {B.}~\bibnamefont {{Crinquand}}},\ }\href {https://doi.org/10.1051/0004-6361/202346781} {\bibfield  {journal} {\bibinfo  {journal} {\aap}\ }\textbf {\bibinfo {volume} {677}},\ \bibinfo {eid} {A67} (\bibinfo {year} {2023})},\ \Eprint {https://arxiv.org/abs/2305.01689} {arXiv:2305.01689 [astro-ph.HE]} \BibitemShut {NoStop}%
\bibitem [{\citenamefont {Yuan}\ \emph {et~al.}(2020)\citenamefont {Yuan}, \citenamefont {Beloborodov}, \citenamefont {Chen},\ and\ \citenamefont {Levin}}]{yuan_plasmoid_2020}%
  \BibitemOpen
  \bibfield  {author} {\bibinfo {author} {\bibfnamefont {Y.}~\bibnamefont {Yuan}}, \bibinfo {author} {\bibfnamefont {A.~M.}\ \bibnamefont {Beloborodov}}, \bibinfo {author} {\bibfnamefont {A.~Y.}\ \bibnamefont {Chen}},\ and\ \bibinfo {author} {\bibfnamefont {Y.}~\bibnamefont {Levin}},\ }\href {https://doi.org/10.3847/2041-8213/abafa8} {\bibfield  {journal} {\bibinfo  {journal} {The Astrophysical Journal Letters}\ }\textbf {\bibinfo {volume} {900}},\ \bibinfo {pages} {L21} (\bibinfo {year} {2020})}\BibitemShut {NoStop}%
\bibitem [{\citenamefont {Mahlmann}\ \emph {et~al.}(2023)\citenamefont {Mahlmann}, \citenamefont {Philippov}, \citenamefont {Mewes}, \citenamefont {Ripperda}, \citenamefont {Most},\ and\ \citenamefont {Sironi}}]{mahlmann_three-dimensional_2023}%
  \BibitemOpen
  \bibfield  {author} {\bibinfo {author} {\bibfnamefont {J.~F.}\ \bibnamefont {Mahlmann}}, \bibinfo {author} {\bibfnamefont {A.~A.}\ \bibnamefont {Philippov}}, \bibinfo {author} {\bibfnamefont {V.}~\bibnamefont {Mewes}}, \bibinfo {author} {\bibfnamefont {B.}~\bibnamefont {Ripperda}}, \bibinfo {author} {\bibfnamefont {E.~R.}\ \bibnamefont {Most}},\ and\ \bibinfo {author} {\bibfnamefont {L.}~\bibnamefont {Sironi}},\ }\href {https://doi.org/10.3847/2041-8213/accada} {\bibfield  {journal} {\bibinfo  {journal} {The Astrophysical Journal Letters}\ }\textbf {\bibinfo {volume} {947}},\ \bibinfo {pages} {L34} (\bibinfo {year} {2023})}\BibitemShut {NoStop}%
\bibitem [{\citenamefont {{Petschek}}(1964)}]{Petschek1964}%
  \BibitemOpen
  \bibfield  {author} {\bibinfo {author} {\bibfnamefont {H.~E.}\ \bibnamefont {{Petschek}}},\ }in\ \href@noop {} {\emph {\bibinfo {booktitle} {NASA Special Publication}}},\ Vol.~\bibinfo {volume} {50},\ \bibinfo {editor} {edited by\ \bibinfo {editor} {\bibfnamefont {W.~N.}\ \bibnamefont {{Hess}}}}\ (\bibinfo  {publisher} {NASA},\ \bibinfo {year} {1964})\ p.\ \bibinfo {pages} {425}\BibitemShut {NoStop}%
\bibitem [{\citenamefont {Robbins}\ and\ \citenamefont {Spitkovsky}(2025)}]{robbins_transition_2025}%
  \BibitemOpen
  \bibfield  {author} {\bibinfo {author} {\bibfnamefont {A.}~\bibnamefont {Robbins}}\ and\ \bibinfo {author} {\bibfnamefont {A.}~\bibnamefont {Spitkovsky}},\ }\href {https://doi.org/10.3847/1538-4357/ae1618} {\bibfield  {journal} {\bibinfo  {journal} {The Astrophysical Journal}\ }\textbf {\bibinfo {volume} {995}},\ \bibinfo {pages} {163} (\bibinfo {year} {2025})}\BibitemShut {NoStop}%
\bibitem [{\citenamefont {{Shibata}}\ and\ \citenamefont {{Tanuma}}(2001)}]{Shibata2001}%
  \BibitemOpen
  \bibfield  {author} {\bibinfo {author} {\bibfnamefont {K.}~\bibnamefont {{Shibata}}}\ and\ \bibinfo {author} {\bibfnamefont {S.}~\bibnamefont {{Tanuma}}},\ }\href {https://doi.org/10.1186/BF03353258} {\bibfield  {journal} {\bibinfo  {journal} {Earth, Planets and Space}\ }\textbf {\bibinfo {volume} {53}},\ \bibinfo {pages} {473} (\bibinfo {year} {2001})},\ \Eprint {https://arxiv.org/abs/astro-ph/0101008} {arXiv:astro-ph/0101008 [astro-ph]} \BibitemShut {NoStop}%
\bibitem [{\citenamefont {{Uzdensky}}\ \emph {et~al.}(2010)\citenamefont {{Uzdensky}}, \citenamefont {{Loureiro}},\ and\ \citenamefont {{Schekochihin}}}]{Uzdensky2010}%
  \BibitemOpen
  \bibfield  {author} {\bibinfo {author} {\bibfnamefont {D.~A.}\ \bibnamefont {{Uzdensky}}}, \bibinfo {author} {\bibfnamefont {N.~F.}\ \bibnamefont {{Loureiro}}},\ and\ \bibinfo {author} {\bibfnamefont {A.~A.}\ \bibnamefont {{Schekochihin}}},\ }\href {https://doi.org/10.1103/PhysRevLett.105.235002} {\bibfield  {journal} {\bibinfo  {journal} {\prl}\ }\textbf {\bibinfo {volume} {105}},\ \bibinfo {eid} {235002} (\bibinfo {year} {2010})},\ \Eprint {https://arxiv.org/abs/1008.3330} {arXiv:1008.3330 [astro-ph.SR]} \BibitemShut {NoStop}%
\bibitem [{\citenamefont {{Loureiro}}\ \emph {et~al.}(2012)\citenamefont {{Loureiro}}, \citenamefont {{Samtaney}}, \citenamefont {{Schekochihin}},\ and\ \citenamefont {{Uzdensky}}}]{Loureiro2012}%
  \BibitemOpen
  \bibfield  {author} {\bibinfo {author} {\bibfnamefont {N.~F.}\ \bibnamefont {{Loureiro}}}, \bibinfo {author} {\bibfnamefont {R.}~\bibnamefont {{Samtaney}}}, \bibinfo {author} {\bibfnamefont {A.~A.}\ \bibnamefont {{Schekochihin}}},\ and\ \bibinfo {author} {\bibfnamefont {D.~A.}\ \bibnamefont {{Uzdensky}}},\ }\href {https://doi.org/10.1063/1.3703318} {\bibfield  {journal} {\bibinfo  {journal} {Physics of Plasmas}\ }\textbf {\bibinfo {volume} {19}},\ \bibinfo {pages} {042303} (\bibinfo {year} {2012})},\ \Eprint {https://arxiv.org/abs/1108.4040} {arXiv:1108.4040 [astro-ph.SR]} \BibitemShut {NoStop}%
\bibitem [{\citenamefont {Sironi}\ \emph {et~al.}(2016)\citenamefont {Sironi}, \citenamefont {Giannios},\ and\ \citenamefont {Petropoulou}}]{sironi_plasmoids_2016}%
  \BibitemOpen
  \bibfield  {author} {\bibinfo {author} {\bibfnamefont {L.}~\bibnamefont {Sironi}}, \bibinfo {author} {\bibfnamefont {D.}~\bibnamefont {Giannios}},\ and\ \bibinfo {author} {\bibfnamefont {M.}~\bibnamefont {Petropoulou}},\ }\href {https://doi.org/10.1093/mnras/stw1620} {\bibfield  {journal} {\bibinfo  {journal} {Monthly Notices of the Royal Astronomical Society}\ }\textbf {\bibinfo {volume} {462}},\ \bibinfo {pages} {48} (\bibinfo {year} {2016})}\BibitemShut {NoStop}%
\bibitem [{\citenamefont {Sironi}\ \emph {et~al.}(2025)\citenamefont {Sironi}, \citenamefont {Uzdensky},\ and\ \citenamefont {Giannios}}]{sironi_relativistic_2025}%
  \BibitemOpen
  \bibfield  {author} {\bibinfo {author} {\bibfnamefont {L.}~\bibnamefont {Sironi}}, \bibinfo {author} {\bibfnamefont {D.~A.}\ \bibnamefont {Uzdensky}},\ and\ \bibinfo {author} {\bibfnamefont {D.}~\bibnamefont {Giannios}},\ }\bibfield  {journal} {\bibinfo  {journal} {Annual Review of Astronomy and Astrophysics}\ }\href {https://doi.org/10.1146/annurev-astro-020325-115713} {10.1146/annurev-astro-020325-115713} (\bibinfo {year} {2025}),\ \bibinfo {note} {arXiv:2506.02101 [astro-ph]}\BibitemShut {NoStop}%
\bibitem [{\citenamefont {{Drake}}\ \emph {et~al.}(2006)\citenamefont {{Drake}}, \citenamefont {{Swisdak}}, \citenamefont {{Che}},\ and\ \citenamefont {{Shay}}}]{Drake2006}%
  \BibitemOpen
  \bibfield  {author} {\bibinfo {author} {\bibfnamefont {J.~F.}\ \bibnamefont {{Drake}}}, \bibinfo {author} {\bibfnamefont {M.}~\bibnamefont {{Swisdak}}}, \bibinfo {author} {\bibfnamefont {H.}~\bibnamefont {{Che}}},\ and\ \bibinfo {author} {\bibfnamefont {M.~A.}\ \bibnamefont {{Shay}}},\ }\href {https://doi.org/10.1038/nature05116} {\bibfield  {journal} {\bibinfo  {journal} {\nat}\ }\textbf {\bibinfo {volume} {443}},\ \bibinfo {pages} {553} (\bibinfo {year} {2006})}\BibitemShut {NoStop}%
\bibitem [{\citenamefont {{Sironi}}\ and\ \citenamefont {{Spitkovsky}}(2014)}]{Sironi2014}%
  \BibitemOpen
  \bibfield  {author} {\bibinfo {author} {\bibfnamefont {L.}~\bibnamefont {{Sironi}}}\ and\ \bibinfo {author} {\bibfnamefont {A.}~\bibnamefont {{Spitkovsky}}},\ }\href {https://doi.org/10.1088/2041-8205/783/1/L21} {\bibfield  {journal} {\bibinfo  {journal} {\apjl}\ }\textbf {\bibinfo {volume} {783}},\ \bibinfo {eid} {L21} (\bibinfo {year} {2014})},\ \Eprint {https://arxiv.org/abs/1401.5471} {arXiv:1401.5471 [astro-ph.HE]} \BibitemShut {NoStop}%
\bibitem [{\citenamefont {{Guo}}\ \emph {et~al.}(2015)\citenamefont {{Guo}}, \citenamefont {{Liu}}, \citenamefont {{Daughton}},\ and\ \citenamefont {{Li}}}]{Guo2015}%
  \BibitemOpen
  \bibfield  {author} {\bibinfo {author} {\bibfnamefont {F.}~\bibnamefont {{Guo}}}, \bibinfo {author} {\bibfnamefont {Y.-H.}\ \bibnamefont {{Liu}}}, \bibinfo {author} {\bibfnamefont {W.}~\bibnamefont {{Daughton}}},\ and\ \bibinfo {author} {\bibfnamefont {H.}~\bibnamefont {{Li}}},\ }\href {https://doi.org/10.1088/0004-637X/806/2/167} {\bibfield  {journal} {\bibinfo  {journal} {\apj}\ }\textbf {\bibinfo {volume} {806}},\ \bibinfo {eid} {167} (\bibinfo {year} {2015})},\ \Eprint {https://arxiv.org/abs/1504.02193} {arXiv:1504.02193 [astro-ph.HE]} \BibitemShut {NoStop}%
\bibitem [{\citenamefont {Petropoulou}\ \emph {et~al.}(2016)\citenamefont {Petropoulou}, \citenamefont {Giannios},\ and\ \citenamefont {Sironi}}]{petropoulou_blazar_2016}%
  \BibitemOpen
  \bibfield  {author} {\bibinfo {author} {\bibfnamefont {M.}~\bibnamefont {Petropoulou}}, \bibinfo {author} {\bibfnamefont {D.}~\bibnamefont {Giannios}},\ and\ \bibinfo {author} {\bibfnamefont {L.}~\bibnamefont {Sironi}},\ }\href {https://doi.org/10.1093/mnras/stw1832} {\bibfield  {journal} {\bibinfo  {journal} {Monthly Notices of the Royal Astronomical Society}\ }\textbf {\bibinfo {volume} {462}},\ \bibinfo {pages} {3325} (\bibinfo {year} {2016})}\BibitemShut {NoStop}%
\bibitem [{\citenamefont {{Cerutti}}\ \emph {et~al.}(2012)\citenamefont {{Cerutti}}, \citenamefont {{Werner}}, \citenamefont {{Uzdensky}},\ and\ \citenamefont {{Begelman}}}]{Cerutti2012}%
  \BibitemOpen
  \bibfield  {author} {\bibinfo {author} {\bibfnamefont {B.}~\bibnamefont {{Cerutti}}}, \bibinfo {author} {\bibfnamefont {G.~R.}\ \bibnamefont {{Werner}}}, \bibinfo {author} {\bibfnamefont {D.~A.}\ \bibnamefont {{Uzdensky}}},\ and\ \bibinfo {author} {\bibfnamefont {M.~C.}\ \bibnamefont {{Begelman}}},\ }\href {https://doi.org/10.1088/2041-8205/754/2/L33} {\bibfield  {journal} {\bibinfo  {journal} {\apjl}\ }\textbf {\bibinfo {volume} {754}},\ \bibinfo {eid} {L33} (\bibinfo {year} {2012})},\ \Eprint {https://arxiv.org/abs/1205.3210} {arXiv:1205.3210 [astro-ph.HE]} \BibitemShut {NoStop}%
\bibitem [{\citenamefont {{Hosking}}\ and\ \citenamefont {{Sironi}}(2020)}]{Hosking2020}%
  \BibitemOpen
  \bibfield  {author} {\bibinfo {author} {\bibfnamefont {D.~N.}\ \bibnamefont {{Hosking}}}\ and\ \bibinfo {author} {\bibfnamefont {L.}~\bibnamefont {{Sironi}}},\ }\href {https://doi.org/10.3847/2041-8213/abafa6} {\bibfield  {journal} {\bibinfo  {journal} {\apjl}\ }\textbf {\bibinfo {volume} {900}},\ \bibinfo {eid} {L23} (\bibinfo {year} {2020})},\ \Eprint {https://arxiv.org/abs/2007.14992} {arXiv:2007.14992 [astro-ph.HE]} \BibitemShut {NoStop}%
\bibitem [{\citenamefont {{Lyubarsky}}(2019)}]{Lyubarsky2019}%
  \BibitemOpen
  \bibfield  {author} {\bibinfo {author} {\bibfnamefont {Y.}~\bibnamefont {{Lyubarsky}}},\ }\href {https://doi.org/10.1093/mnras/sty3233} {\bibfield  {journal} {\bibinfo  {journal} {\mnras}\ }\textbf {\bibinfo {volume} {483}},\ \bibinfo {pages} {1731} (\bibinfo {year} {2019})},\ \Eprint {https://arxiv.org/abs/1811.11122} {arXiv:1811.11122 [astro-ph.HE]} \BibitemShut {NoStop}%
\bibitem [{\citenamefont {{Lyubarsky}}(2020)}]{Yyubarsky2020}%
  \BibitemOpen
  \bibfield  {author} {\bibinfo {author} {\bibfnamefont {Y.}~\bibnamefont {{Lyubarsky}}},\ }\href {https://doi.org/10.3847/1538-4357/ab97b5} {\bibfield  {journal} {\bibinfo  {journal} {\apj}\ }\textbf {\bibinfo {volume} {897}},\ \bibinfo {eid} {1} (\bibinfo {year} {2020})},\ \Eprint {https://arxiv.org/abs/2001.02007} {arXiv:2001.02007 [astro-ph.HE]} \BibitemShut {NoStop}%
\bibitem [{\citenamefont {Mahlmann}\ \emph {et~al.}(2022)\citenamefont {Mahlmann}, \citenamefont {Philippov}, \citenamefont {Levinson}, \citenamefont {Spitkovsky},\ and\ \citenamefont {Hakobyan}}]{mahlmann_electromagnetic_2022}%
  \BibitemOpen
  \bibfield  {author} {\bibinfo {author} {\bibfnamefont {J.~F.}\ \bibnamefont {Mahlmann}}, \bibinfo {author} {\bibfnamefont {A.~A.}\ \bibnamefont {Philippov}}, \bibinfo {author} {\bibfnamefont {A.}~\bibnamefont {Levinson}}, \bibinfo {author} {\bibfnamefont {A.}~\bibnamefont {Spitkovsky}},\ and\ \bibinfo {author} {\bibfnamefont {H.}~\bibnamefont {Hakobyan}},\ }\href {https://doi.org/10.3847/2041-8213/ac7156} {\bibfield  {journal} {\bibinfo  {journal} {The Astrophysical Journal Letters}\ }\textbf {\bibinfo {volume} {932}},\ \bibinfo {pages} {L20} (\bibinfo {year} {2022})}\BibitemShut {NoStop}%
\bibitem [{\citenamefont {{Zelenyi}}\ and\ \citenamefont {{Krasnoselskikh}}(1979)}]{Zelenzi1979}%
  \BibitemOpen
  \bibfield  {author} {\bibinfo {author} {\bibfnamefont {L.~M.}\ \bibnamefont {{Zelenyi}}}\ and\ \bibinfo {author} {\bibfnamefont {V.~V.}\ \bibnamefont {{Krasnoselskikh}}},\ }\href@noop {} {\bibfield  {journal} {\bibinfo  {journal} {\sovast}\ }\textbf {\bibinfo {volume} {23}},\ \bibinfo {pages} {460} (\bibinfo {year} {1979})}\BibitemShut {NoStop}%
\bibitem [{\citenamefont {Hoshino}(2020)}]{hoshino_stabilization_2020}%
  \BibitemOpen
  \bibfield  {author} {\bibinfo {author} {\bibfnamefont {M.}~\bibnamefont {Hoshino}},\ }\href {https://doi.org/10.3847/1538-4357/aba59d} {\bibfield  {journal} {\bibinfo  {journal} {The Astrophysical Journal}\ }\textbf {\bibinfo {volume} {900}},\ \bibinfo {pages} {66} (\bibinfo {year} {2020})}\BibitemShut {NoStop}%
\bibitem [{\citenamefont {{Demidov}}\ and\ \citenamefont {{Lyubarsky}}(2025)}]{Demidov2025}%
  \BibitemOpen
  \bibfield  {author} {\bibinfo {author} {\bibfnamefont {I.}~\bibnamefont {{Demidov}}}\ and\ \bibinfo {author} {\bibfnamefont {Y.}~\bibnamefont {{Lyubarsky}}},\ }\href {https://doi.org/10.3847/1538-4357/ad9d3a} {\bibfield  {journal} {\bibinfo  {journal} {\apj}\ }\textbf {\bibinfo {volume} {979}},\ \bibinfo {eid} {104} (\bibinfo {year} {2025})},\ \Eprint {https://arxiv.org/abs/2410.16167} {arXiv:2410.16167 [astro-ph.HE]} \BibitemShut {NoStop}%
\bibitem [{\citenamefont {{Fried}}(1959)}]{Fried1959}%
  \BibitemOpen
  \bibfield  {author} {\bibinfo {author} {\bibfnamefont {B.~D.}\ \bibnamefont {{Fried}}},\ }\href {https://doi.org/10.1063/1.1705933} {\bibfield  {journal} {\bibinfo  {journal} {Physics of Fluids}\ }\textbf {\bibinfo {volume} {2}},\ \bibinfo {pages} {337} (\bibinfo {year} {1959})}\BibitemShut {NoStop}%
\bibitem [{\citenamefont {{Weibel}}(1959)}]{Weibel1959}%
  \BibitemOpen
  \bibfield  {author} {\bibinfo {author} {\bibfnamefont {E.~S.}\ \bibnamefont {{Weibel}}},\ }\href {https://doi.org/10.1103/PhysRevLett.2.83} {\bibfield  {journal} {\bibinfo  {journal} {\prl}\ }\textbf {\bibinfo {volume} {2}},\ \bibinfo {pages} {83} (\bibinfo {year} {1959})}\BibitemShut {NoStop}%
\bibitem [{\citenamefont {Hakobyan}\ \emph {et~al.}(2023)\citenamefont {Hakobyan}, \citenamefont {Spitkovsky}, \citenamefont {Chernoglazov}, \citenamefont {Philippov}, \citenamefont {Groselj},\ and\ \citenamefont {Mahlmann}}]{hayk_hakobyan_2023_7566725}%
  \BibitemOpen
  \bibfield  {author} {\bibinfo {author} {\bibfnamefont {H.}~\bibnamefont {Hakobyan}}, \bibinfo {author} {\bibfnamefont {A.}~\bibnamefont {Spitkovsky}}, \bibinfo {author} {\bibfnamefont {A.}~\bibnamefont {Chernoglazov}}, \bibinfo {author} {\bibfnamefont {A.}~\bibnamefont {Philippov}}, \bibinfo {author} {\bibfnamefont {D.}~\bibnamefont {Groselj}},\ and\ \bibinfo {author} {\bibfnamefont {J.}~\bibnamefont {Mahlmann}},\ }\href {https://doi.org/10.5281/zenodo.7566725} {\bibinfo {title} {Princetonuniversity/tristan-mp-v2: v2.6}} (\bibinfo {year} {2023})\BibitemShut {NoStop}%
\bibitem [{\citenamefont {{Uzdensky}}\ \emph {et~al.}(2011)\citenamefont {{Uzdensky}}, \citenamefont {{Cerutti}},\ and\ \citenamefont {{Begelman}}}]{Uzensky2011}%
  \BibitemOpen
  \bibfield  {author} {\bibinfo {author} {\bibfnamefont {D.~A.}\ \bibnamefont {{Uzdensky}}}, \bibinfo {author} {\bibfnamefont {B.}~\bibnamefont {{Cerutti}}},\ and\ \bibinfo {author} {\bibfnamefont {M.~C.}\ \bibnamefont {{Begelman}}},\ }\href {https://doi.org/10.1088/2041-8205/737/2/L40} {\bibfield  {journal} {\bibinfo  {journal} {\apjl}\ }\textbf {\bibinfo {volume} {737}},\ \bibinfo {eid} {L40} (\bibinfo {year} {2011})},\ \Eprint {https://arxiv.org/abs/1105.0942} {arXiv:1105.0942 [astro-ph.HE]} \BibitemShut {NoStop}%
\bibitem [{\citenamefont {Zenitani}\ and\ \citenamefont {Hoshino}(2001)}]{Zenitani2001ThePlasmas}%
  \BibitemOpen
  \bibfield  {author} {\bibinfo {author} {\bibfnamefont {S.}~\bibnamefont {Zenitani}}\ and\ \bibinfo {author} {\bibfnamefont {M.}~\bibnamefont {Hoshino}},\ }\href {https://doi.org/10.1086/337972} {\bibfield  {journal} {\bibinfo  {journal} {The Astrophysical Journal}\ }\textbf {\bibinfo {volume} {562}},\ \bibinfo {pages} {L63} (\bibinfo {year} {2001})}\BibitemShut {NoStop}%
\bibitem [{\citenamefont {Goodbred}\ and\ \citenamefont {Liu}(2022)}]{goodbred_first-principles_2022}%
  \BibitemOpen
  \bibfield  {author} {\bibinfo {author} {\bibfnamefont {M.}~\bibnamefont {Goodbred}}\ and\ \bibinfo {author} {\bibfnamefont {Y.-H.}\ \bibnamefont {Liu}},\ }\href {https://doi.org/10.1103/PhysRevLett.129.265101} {\bibfield  {journal} {\bibinfo  {journal} {Physical Review Letters}\ }\textbf {\bibinfo {volume} {129}},\ \bibinfo {pages} {265101} (\bibinfo {year} {2022})}\BibitemShut {NoStop}%
\bibitem [{\citenamefont {{French}}\ \emph {et~al.}(2023)\citenamefont {{French}}, \citenamefont {{Guo}}, \citenamefont {{Zhang}},\ and\ \citenamefont {{Uzdensky}}}]{French2023}%
  \BibitemOpen
  \bibfield  {author} {\bibinfo {author} {\bibfnamefont {O.}~\bibnamefont {{French}}}, \bibinfo {author} {\bibfnamefont {F.}~\bibnamefont {{Guo}}}, \bibinfo {author} {\bibfnamefont {Q.}~\bibnamefont {{Zhang}}},\ and\ \bibinfo {author} {\bibfnamefont {D.~A.}\ \bibnamefont {{Uzdensky}}},\ }\href {https://doi.org/10.3847/1538-4357/acb7dd} {\bibfield  {journal} {\bibinfo  {journal} {\apj}\ }\textbf {\bibinfo {volume} {948}},\ \bibinfo {eid} {19} (\bibinfo {year} {2023})},\ \Eprint {https://arxiv.org/abs/2210.08358} {arXiv:2210.08358 [astro-ph.HE]} \BibitemShut {NoStop}%
\bibitem [{\citenamefont {Liu}\ \emph {et~al.}(2017)\citenamefont {Liu}, \citenamefont {Hesse}, \citenamefont {Guo}, \citenamefont {Daughton}, \citenamefont {Li}, \citenamefont {Cassak},\ and\ \citenamefont {Shay}}]{liu_why_2017}%
  \BibitemOpen
  \bibfield  {author} {\bibinfo {author} {\bibfnamefont {Y.-H.}\ \bibnamefont {Liu}}, \bibinfo {author} {\bibfnamefont {M.}~\bibnamefont {Hesse}}, \bibinfo {author} {\bibfnamefont {F.}~\bibnamefont {Guo}}, \bibinfo {author} {\bibfnamefont {W.}~\bibnamefont {Daughton}}, \bibinfo {author} {\bibfnamefont {H.}~\bibnamefont {Li}}, \bibinfo {author} {\bibfnamefont {P.}~\bibnamefont {Cassak}},\ and\ \bibinfo {author} {\bibfnamefont {M.}~\bibnamefont {Shay}},\ }\href {https://doi.org/10.1103/PhysRevLett.118.085101} {\bibfield  {journal} {\bibinfo  {journal} {Physical Review Letters}\ }\textbf {\bibinfo {volume} {118}},\ \bibinfo {pages} {085101} (\bibinfo {year} {2017})}\BibitemShut {NoStop}%
\bibitem [{\citenamefont {{Philippov}}\ \emph {et~al.}(2019)\citenamefont {{Philippov}}, \citenamefont {{Uzdensky}}, \citenamefont {{Spitkovsky}},\ and\ \citenamefont {{Cerutti}}}]{Philippov2019_nanoshots}%
  \BibitemOpen
  \bibfield  {author} {\bibinfo {author} {\bibfnamefont {A.}~\bibnamefont {{Philippov}}}, \bibinfo {author} {\bibfnamefont {D.~A.}\ \bibnamefont {{Uzdensky}}}, \bibinfo {author} {\bibfnamefont {A.}~\bibnamefont {{Spitkovsky}}},\ and\ \bibinfo {author} {\bibfnamefont {B.}~\bibnamefont {{Cerutti}}},\ }\href {https://doi.org/10.3847/2041-8213/ab1590} {\bibfield  {journal} {\bibinfo  {journal} {\apjl}\ }\textbf {\bibinfo {volume} {876}},\ \bibinfo {eid} {L6} (\bibinfo {year} {2019})},\ \Eprint {https://arxiv.org/abs/1902.07730} {arXiv:1902.07730 [astro-ph.HE]} \BibitemShut {NoStop}%
\bibitem [{\citenamefont {Ball}\ \emph {et~al.}(2019)\citenamefont {Ball}, \citenamefont {Sironi},\ and\ \citenamefont {Özel}}]{ball_mechanism_2019}%
  \BibitemOpen
  \bibfield  {author} {\bibinfo {author} {\bibfnamefont {D.}~\bibnamefont {Ball}}, \bibinfo {author} {\bibfnamefont {L.}~\bibnamefont {Sironi}},\ and\ \bibinfo {author} {\bibfnamefont {F.}~\bibnamefont {Özel}},\ }\href {https://doi.org/10.3847/1538-4357/ab3f2e} {\bibfield  {journal} {\bibinfo  {journal} {The Astrophysical Journal}\ }\textbf {\bibinfo {volume} {884}},\ \bibinfo {pages} {57} (\bibinfo {year} {2019})}\BibitemShut {NoStop}%
\bibitem [{\citenamefont {Rowan}\ \emph {et~al.}(2019)\citenamefont {Rowan}, \citenamefont {Sironi},\ and\ \citenamefont {Narayan}}]{rowan_electron_2019}%
  \BibitemOpen
  \bibfield  {author} {\bibinfo {author} {\bibfnamefont {M.~E.}\ \bibnamefont {Rowan}}, \bibinfo {author} {\bibfnamefont {L.}~\bibnamefont {Sironi}},\ and\ \bibinfo {author} {\bibfnamefont {R.}~\bibnamefont {Narayan}},\ }\href {https://doi.org/10.3847/1538-4357/ab03d7} {\bibfield  {journal} {\bibinfo  {journal} {The Astrophysical Journal}\ }\textbf {\bibinfo {volume} {873}},\ \bibinfo {pages} {2} (\bibinfo {year} {2019})}\BibitemShut {NoStop}%
\bibitem [{\citenamefont {Zhang}\ \emph {et~al.}(2020)\citenamefont {Zhang}, \citenamefont {Li}, \citenamefont {Giannios}, \citenamefont {Guo}, \citenamefont {Liu},\ and\ \citenamefont {Dong}}]{zhang_radiation_2020}%
  \BibitemOpen
  \bibfield  {author} {\bibinfo {author} {\bibfnamefont {H.}~\bibnamefont {Zhang}}, \bibinfo {author} {\bibfnamefont {X.}~\bibnamefont {Li}}, \bibinfo {author} {\bibfnamefont {D.}~\bibnamefont {Giannios}}, \bibinfo {author} {\bibfnamefont {F.}~\bibnamefont {Guo}}, \bibinfo {author} {\bibfnamefont {Y.-H.}\ \bibnamefont {Liu}},\ and\ \bibinfo {author} {\bibfnamefont {L.}~\bibnamefont {Dong}},\ }\href {https://doi.org/10.3847/1538-4357/abb1b0} {\bibfield  {journal} {\bibinfo  {journal} {The Astrophysical Journal}\ }\textbf {\bibinfo {volume} {901}},\ \bibinfo {pages} {149} (\bibinfo {year} {2020})}\BibitemShut {NoStop}%
\end{thebibliography}%

\end{document}